\documentclass[runningheads]{llncs}
\usepackage{amsmath}
\usepackage{subcaption}
\usepackage{comment}
\usepackage{booktabs}
\usepackage{makecell} 
\usepackage{amssymb}  
\usepackage{times}

\usepackage[T1]{fontenc}
\usepackage{graphicx}
\usepackage{xcolor}

\newcommand{\demoName}{CityHood}

\begin{document}

\title{
\demoName: An Explainable Travel Recommender System for Cities and Neighborhoods}
\titlerunning{\demoName}
%
\author{Gustavo H. Santos\inst{1} \and
Myriam Delgado\inst{1}\and Daniel Silver\inst{2}\and Thiago H Silva\inst{1,2}}
\authorrunning{Santos et al.}

\institute{Universidade Tecnológica Federal do Paraná, Curitiba, Brazil \and
University of Toronto, Toronto, Canada
\email{gustavohenriquesantos@alunos.utfpr.edu.br, \{myriamdelg,thiagoh\}@utfpr.edu.br, dan.silver@utoronto.ca}}
\maketitle              
\begin{abstract}
We present \demoName, an interactive and explainable recommendation system that suggests cities and neighborhoods based on users' areas of interest. The system models user interests leveraging large-scale Google Places reviews enriched with geographic, socio-demographic, political, and cultural indicators. It provides personalized recommendations at city (Core-Based Statistical Areas - CBSAs) and neighborhood (ZIP code) levels, supported by an explainable technique (LIME) and natural-language explanations. Users can explore recommendations based on their stated preferences and inspect the reasoning behind each suggestion through a visual interface. The demo illustrates how spatial similarity, cultural alignment, and interest understanding can be used to make travel recommendations transparent and engaging. This work bridges gaps in location-based recommendation by combining a kind of interest modeling, multi-scale analysis, and explainability in a user-facing system.

\end{abstract}

\section{Introduction}

Understanding individual mobility preferences is key to building smarter urban services, from tourism platforms to personalized navigation~\cite{chen2024_HumanMobility}. 
While existing location-based recommenders often suggest points of interest (POIs), they typically overlook why users prefer certain cities or neighborhoods, and rarely provide explanations for their choices. This lack of interpretability limits user trust and system insight~\cite{ranjbar2024explaining}.

To address this, we present {\demoName}\footnote{\url{https://cityhood.vercel.app/}}, an interactive, human-interpretable recommendation system that models user interests at the city (Core-Based Statistical Areas - CBSAs), and neighborhood (ZIP code) levels, recommending new destinations by comparing socio-demographic, political, geographic, and cultural features of previously visited locations. The system is enhanced with an explainable technique (local interpretable model-agnostic explanations - LIME~\cite{lime}) and natural language explanations.

Our demo allows users to: (i) Select places they liked or disliked (cities or neighborhoods); (ii) Receive transparent, historic-based travel recommendations; (iii) Explore feature-level similarities and debug model outputs.

The system bridges gaps in urban recommendations by combining spatial granularity (cities and neighborhoods), interest modeling, and explainability in a user-facing application. Unlike black-box POI recommenders, it provides actionable insights into why a place may appeal to a particular user based on their past preferences.

\section{Related Work}
\label{sec:related_works}

\paragraph{Human Mobility Modeling}  
Understanding urban mobility has evolved through studies on visitation patterns~\cite{schlapfer2021universal,xu2021emergence} and tourist clustering~\cite{predictour}. Recent work emphasizes the importance of contextual and behavioral diversity~\cite{napoli2024one} and the need for transparency in AI-driven mobility systems~\cite{chen2024_HumanMobility,liang2024exploring}. Inspired by~\cite{Santos_Gubert_Delgado_Silva_2025}, our system models preferences across geographic, political, socio-economic, and cultural features, and extends that approach by using explainable AI, an aspect not addressed in the original work.
\paragraph{Place Recommendation Systems}  
Most place recommenders focus on point-of-interest (POI) suggestions~\cite{cheng2012fused,10035007,8848605,9036967}, with limited attention to city- or neighborhood-level granularity. Notable exceptions include approaches using spatial embeddings~\cite{cheng2012fused} and hierarchical context models~\cite{8848605}. However, few incorporate behavioral traits or offer interpretability. Our demo addresses this gap through an explainable, multi-level recommendation system that predicts user interest at city and neighborhood scales.
\paragraph{Explainability in Recommenders}  
While explainable recommendations have gained traction~\cite{de2024explainable,ranjbar2024explaining}, it remains rare in mobility-focused systems. Our demo uniquely integrates local model explanations (LIME~\cite{lime}) with users' interest per areas segmentation to provide transparent, user-facing justifications.

\section{Methodology}
\label{sec:methodology}

\subsection{Dataset}
\label{sec:Dataset}

We use a large-scale dataset from Google Places~\cite{li_uctopic_2022,10.1145/3539618.3592036}, comprising over 666 million geo-tagged reviews on nearly 5 million U.S. venues, written by 113 million users. To capture tourism-oriented behavior, we filtered for users who reviewed venues in at least six different Core-Based Statistical Areas (CBSAs) —urban-centric regions defined by the U.S. Census Bureau—resulting in 245 million reviews by 4.6 million users.

CBSAs serve as our city-level units, while ZIP codes represent neighborhood-level granularity. User interest in a particular area is proxied by review volume, and both CBSAs and ZIP codes are ranked, per user, using dense ranking (ties share rank, next rank is sequential). The review patterns across top-ranked CBSAs show skewed activity distributions typical of location-based platforms.

To contextualize recommendations, each region is enriched with socio-demographic attributes (e.g., race, income, education, employment) from the NHGIS database~\cite{NHGIS}, political leaning from 2020 U.S. election data, and cultural profiles using Scenes Theory~\cite{culture_fingerprint}. ZIP and CBSA boundaries are sourced from the U.S. Census TIGER database.

The enriched data enable our demo to offer context-sensitive recommendations across cities and neighborhoods, informed by user mobility and regional characteristics.

\subsection{Model Design and Training}

Our recommendation system learns which types of cities and neighborhoods users are most likely to prefer based on their past activity. We assume that users show stronger interest in regions where they have written more reviews. Using this assumption, we divide user’s visited regions into: \textbf{Top regions:} Cities (CBSAs) or neighborhoods (ZIP codes) with the highest number of reviews; and \textbf{Bottom regions:} All other visited locations with lower review counts.

At the city level, we select the user’s top-$k$ cities. At the neighborhood level, we identify the top-$m$ ZIP codes within those cities. This enables a fine-grained view of user preferences across both broader urban areas and specific local neighborhoods.

To recommend each region, first, we calculate how similar it is to the user's past top and bottom regions across multiple dimensions, including geographic distance, population, income, education level, racial composition, political leaning, cultural traits (via Scenes Theory), and types of venues. These (dis)similarities are then aggregated into a feature vector that describes the region based on the user's past preferences.

Using these features, we train a LightGBM binary classifier to predict whether a region is likely to be of high interest to the user. The model is trained separately for cities and neighborhoods. To support explainability, we apply LIME to explain each prediction and an LLM to provide natural-language summaries. The demo interface displays these explanations to help users understand each recommendation.

Using an 80/20 train-test split, our LightGBM model demonstrated strong performance in city-level recommendations, achieving recall scores between 0.66 and 0.79, and F1-scores between 0.56 and 0.65 for values of $k$ ranging from 2 to 5. These results consistently outperformed both the popularity-based and item-based collaborative filtering (ICF) baselines. For instance, at $k = 2$, our model reached a recall of 0.79, compared to 0.31 and 0.63 for the popularity and ICF baselines, respectively. Similar trends were observed at the neighborhood level, where the model achieved recall between 0.59 and 0.77, and F1-scores from 0.47 to 0.75—further widening the performance gap over the baselines.
 
\section{The {\demoName} Framework} 
\label{sec:demo}

To demonstrate our city and neighborhood recommendation framework ({\demoName})  publicly accessible at \underline{\url{https://cityhood.vercel.app/}}, we have developed an interactive web application. The system guides users through a two-stage process: \textbf{city-level recommendations} followed by \textbf{neighborhood-level refinement}, providing explainable feedback at each stage, supported by visual and textual justifications. The interface emphasizes 
explainability and engagement, enabling users to interrogate recommendations through feature-level insights, natural-language explanations, and curated visual cues.

\vspace{-0.4cm}
\subsection{City-Level Recommendation}

\begin{figure}[t]
  \centering
  \begin{subfigure}{0.38\linewidth}
    \centering
    \includegraphics[width=\linewidth]{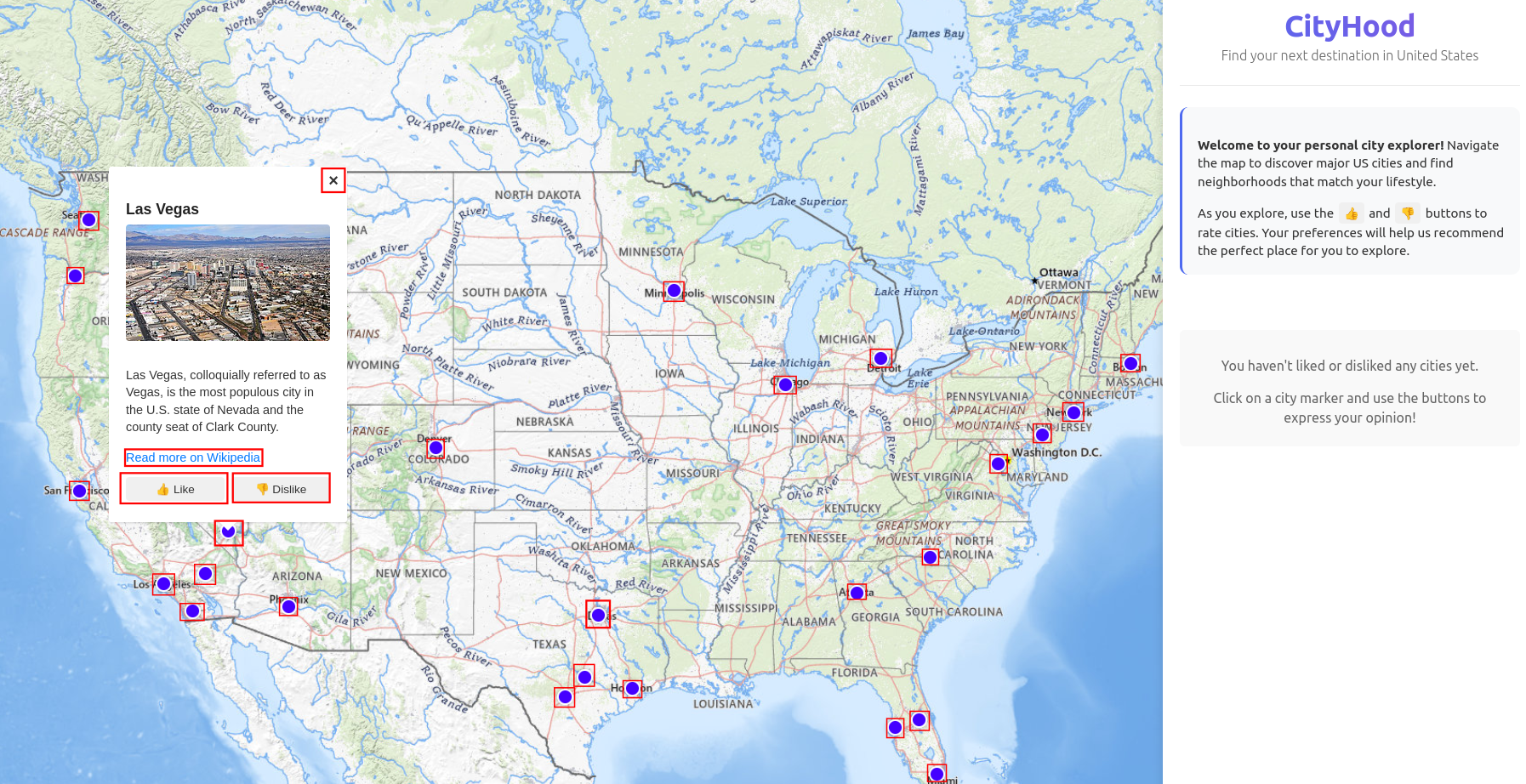}
    \caption{Home Screen}
    \label{fig:home_screen}
  \end{subfigure}%
  \begin{subfigure}{0.30\linewidth}
    \centering
    \includegraphics[width=\linewidth]{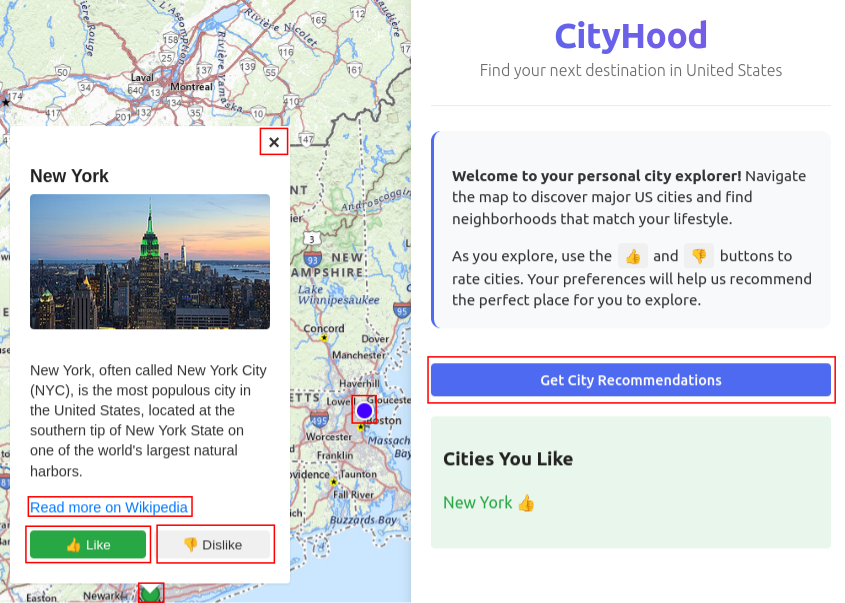}
    \caption{Exploring cities}
    \label{fig:card}
  \end{subfigure}%
  \begin{subfigure}{0.30\linewidth}
    \centering
    \includegraphics[width=\linewidth]{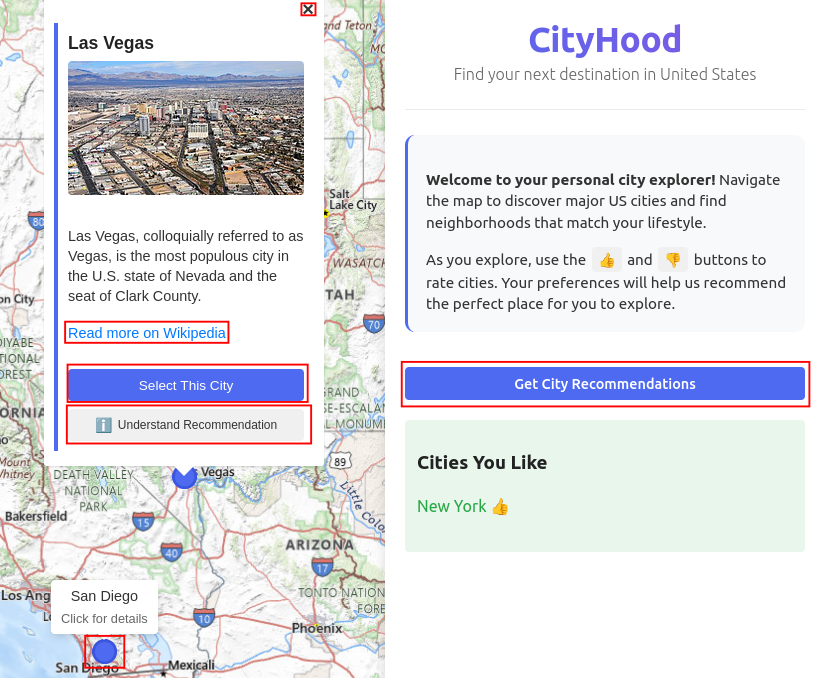}
    \caption{Recommended Cities}
    \label{fig:recommendation_cities}
  \end{subfigure}%
  \caption{City-Level Recommendation}
  \label{fig:city_level_interface}
\end{figure}

The workflow begins with a map interface displaying the top 25 most popular U.S. Core-Based Statistical Areas (Figure~\ref{fig:city_level_interface} - clickable elements are show with red borders). Designed for ease of use, the interface supports quick navigation, allowing users to explore cities through their markers (blue circles in Figure~\ref{fig:city_level_interface}a), which trigger pop-ups featuring a Wikipedia-sourced description and an illustrative image to aid decision-making (e.g., Las Vegas). Cities can then be labeled as “liked” or “disliked” based on user preferences. To receive recommendations, users must label a minimum of one (NY is 'liked' in Figure~\ref{fig:city_level_interface}b) and a maximum of six cities.

Upon submission, the frontend, implemented with React\footnote{\url{https://react.dev/}} and Maplibre\footnote{\url{https://visgl.github.io/react-map-gl/}} for responsive rendering, transmits user feedback to a Dockerized FastAPI\footnote{\url{https://fastapi.tiangolo.com/}} backend. The server processes this input and returns, by default, 3 recommended cities (Fig \ref{fig:recommendation_cities}),from which the user must choose one to have its recommendation explained and to continue to neighborhood-level refinement (Sec \ref{sec:neighoorhood}).
Each recommendation includes: i) A natural-language explanation justifying the recommendation, based on a prompt pre-filled with LIME feature contributions and raw similarity distances, and a button to redirect the user to ChatGPT page with the prompt ready to use; ii) Quantitative metrics, including LIME feature weights and raw similarity scores between the candidate city and the user’s liked/disliked selections. The interface allows users to toggle a checkbox, exposing these values and ensuring transparency.

\subsection{Neighborhood-Level Refinement} 
\label{sec:neighoorhood}

\begin{figure}[t]
  \centering
  \begin{subfigure}{0.45\linewidth}
    \centering
    \includegraphics[width=\linewidth]{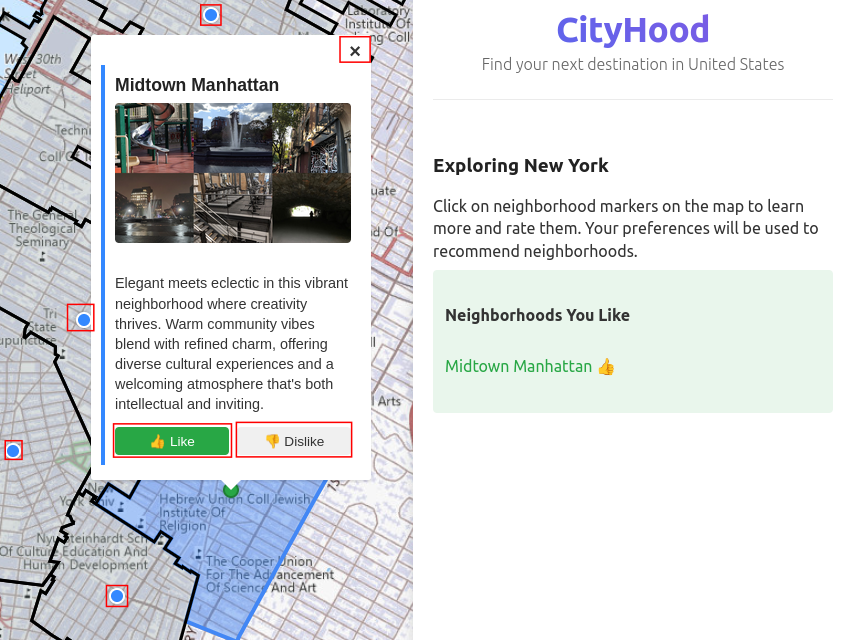}
    \caption{Exploring previous liked cities}
    \label{fig:card}
  \end{subfigure}%
    \begin{subfigure}{0.45\linewidth}
    \centering
    \includegraphics[width=\linewidth]{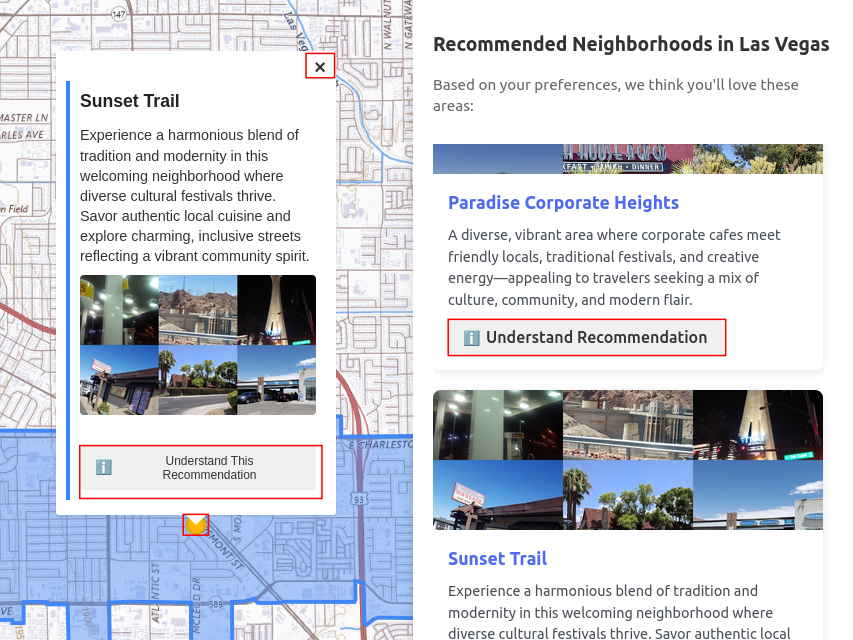}
    \caption{Recommended neighborhoods}
    \label{fig:card}
  \end{subfigure}%
  \caption{Neighborhood-Level Refinement}
  \label{fig:neighborhood_level_interface}
\end{figure}

In the second recommendation level, the system transitions to neighborhood-level refinement (Figure~\ref{fig:neighborhood_level_interface} - clickable elements are show in red borders). Here, for each \textbf{previously liked city} 
users are presented with \textbf{its 10 most popular} ZIP codes (\textbf{neighborhoods}) and must point preferences by labeling them as “liked” or “disliked.”

To help identify neighborhoods, a generated text description and a curated visual preview are provided. Our custom neighborhood characterization solves the lack of structured ZIP code descriptions by using a two-step process to generate dynamic summaries: (1) extracting regional features (the ones used by the model), and (2) using a large language model (specifically, a local deepseek-r1:32b) to generate coherent, human-readable descriptions with those features, through a prompt filled with the values and an explanation of what they mean. 

Visual context is curated through a hybrid strategy. First, candidate images, taken from the reviews of the area, are filtered using the Places365 scene classifier~\cite{NIPS2014_19ea3982}, which categorizes images by place type (e.g., "beach," "forest," and "office"). We prioritize outdoor-related classes to better reflect neighborhood characteristics. Next, a lightweight YOLOv8 nano model refines the selection by filtering out images where the primary subjects are people (e.g., selfies) or vehicles, which may have passed the initial filter due to relevant backgrounds but do not accurately represent the broader environment.

After labeling neighborhood preferences in previously visited cities (e.g., Midtown Manhattan in NY), the user is presented with recommendations in the selected destination city (e.g., Las Vegas). Similarly to the city recommendation, for each recommended neighborhood, the system provides i) A natural-language summary derived from LIME feature weights; ii) Raw similarity/distance metrics and LIME weights for transparency.

\section{Conclusion}
\label{sec:conclusion}

In this work, we presented {\demoName}, an interactive, explainable recommendation system that suggests cities and neighborhoods based on user preferences and regional characteristics. The demo highlights a multi-level approach, operating at both CBSA and ZIP code levels, while integrating spatial, cultural, demographic, and political features. Through natural-language explanations powered by LIME, Wikipedia-based context, and a user-friendly interface, the system enables transparent and engaging travel recommendations. Beyond delivering recommendations, the CityHood framework emphasizes interpretability and user agency. By allowing users to understand why a city or neighborhood is suggested, it supports a more informed and personalized exploration of urban areas. In future iterations, we plan to expand geographic coverage, improve neighborhood characterization using richer review-based NLP techniques, and integrate time-aware (duration, timing) and intent-based factors (e.g., leisure vs. business travel) and explore different LBSNs for less biased recommendations. 

\scriptsize{ \subsubsection{Acknowledgements} This research is partially supported by the SocialNet project (process 2023/00148-0 of FAPESP), by CNPq (proc. 314603/2023-9, 441444/2023-7, 409669/2024-5, and 444724/2024-9) and the INCTs ICoNIoT and TILD-IAR funded by CNPq (proc. 405940/2022-0 and 408490/2024-1) and CAPES FinanceCode 88887.954253/2024-00. }

\vspace{-0.5cm}

{\small \bibliographystyle{splncs04}
\bibliography{bibliography}}
\end{document}